\documentclass[twocolumn,showpacs,preprintnumbers,amsmath,amssymb]{revtex4}

\usepackage{graphicx}
\usepackage{dcolumn}
\usepackage{bm}
\newcommand{\YSO}{Y$_2$SiO$_5$}
\newcommand{\PrYSO}{$Pr^{3+}$:$Y_2SiO_5$}
\renewcommand{\Pr}{praseodymium}

\newcommand{\bB}{\bm{B}}
\newcommand{\bI}{\bm{I}}

\begin{document}
\preprint{APS/123-QED}

\title{Method of extending hyperfine coherence times in $\bm{Pr^{3+}:Y_2SiO_5}$}

\author{E. Fraval}
\email{elliot.fraval@anu.edu.au}
\author{M. J. Sellars}
\author{J. J. Longdell}

\affiliation{Laser Physics Center, Australian National University.}

\date{\today}

\begin{abstract}
  
  In this letter we present a method for increasing the coherence time
  of \Pr\ hyperfine ground state transitions in \PrYSO\ by the
  application of a specific external magnetic field. The magnitude and
  angle of the external field is applied such that the Zeeman
  splitting of a hyperfine transition is at a critical point in three
  dimensions, making the first order Zeeman shift vanishingly small
  for the transition.  This reduces the influence of the magnetic
  interactions between the \Pr\ ions and the spins in the host lattice
  on the transition frequency. Using this method a phase memory time
  of 82ms was observed, a value two orders of magnitude greater than
  previously reported.  It is shown that the residual dephasing is
  amenable quantum error correction.

\end{abstract}

\maketitle

There is growing interest in the use of nuclear spin states associated
with dopant ions in a solid state host to store and manipulate quantum
information\cite{ichi01,kane98,ohls02}. These applications require the
relevant spin transitions to have long coherence times. This can be
challenging to achieve in a solid state system due to magnetic
interactions between the dopant ion and spins within the host. A spin
free host can be utilized to minimize these interactions\cite{kane98},
however not all dopant species of interest are chemically compatible
with such hosts. An example of a class of dopants where no
satisfactory spin free host has been identified, are the rare earth
ions. Because of the potential to manipulate their spin states
optically, the use of rare earth ions has been proposed in a number of
quantum information applications\cite{ichi01,ohls02}. In this letter
we investigate the decoherence in the ground state hyperfine
transitions in \PrYSO\ and demonstrate a new method for increasing
their coherence times.  This technique is expected to be applicable to
a wide range of spin systems.

\PrYSO\ was chosen to investigate since along with its use in a
quantum computing architecture having been proposed by Ichimura et
al.\cite{ichi01}, it has already been used in slow and stopped light
demonstrations\cite{turu02,yama98}. The longest coherence time ($T_2$)
stated in the literature for a ground state hyperfine transition in
\PrYSO\ is $500\mu$s\cite{ham98a}.  This is likely to be insufficient
for practical quantum computing applications given that the hyperfine
transition frequency is $\sim$$10MHz$, limiting the Rabi frequencies
of any driving field to less than a few MHz so as to be transition
specific. This only allows of the order of $1000$ operations to be
completed within $T_2$.

Yttrium orthosilicate has symmetry given by the $C^6_{2h}$ space group
\cite{buer69} with two formula units \YSO\ per translational unit.
This gives four different sites at which the Pr can substitute for Y.
The four sites can be divided into two pairs.  The two pairs labelled
'site 1' and 'site 2' \cite{equa95} have different crystal field
splittings and hence different optical and hyperfine frequencies. The
members of each pair, labelled 'a' and 'b', are related to each other
by the crystal's $C_2$ axis and their optical and hyperfine
frequencies are degenerate in zero field.  The Pr sites in \YSO\ have
electronic singlet ground states with hyperfine ground state
interactions described by the following Hamiltonian:

\begin{equation}
  \label{eq:ZeemanHamiltonian}
  H = \bm{B \cdot M \cdot I+I \cdot Q \cdot I}
\end{equation}

where $\bB$ is the magnetic field vector, $\bI$ is the vector of
nuclear spin operators, $\bm{M}$ is the effective Zeeman tensor
combining nuclear and electronic Zeeman interactions and $\bm{Q}$ is
the effective quadrupole tensor combining the quadrupole and second
order magnetic hyperfine interaction, known as the
pseudo-quadrupole\cite{bake58}. The one naturally occurring isotope of
Pr has a nuclear spin of 5/2. $\bm{M}$ and $\bm{Q}$ for site 1 were
recently determined for \PrYSO\ to be:

\begin{eqnarray}
  \label{eq:QuadTensor}
  \bm{Q} & = & R(\alpha,\beta,\gamma)
  \begin{bmatrix}
    -E&0&0\\0&E&0\\0&0&D
  \end{bmatrix}
 R^T(\alpha,\beta,\gamma)\nonumber\\
    \bm{M} & = & R(\alpha,\beta,\gamma)
  \begin{bmatrix}
    g_x&0&0\\0&g_y&0\\0&0&g_z
  \end{bmatrix}
 R^T(\alpha,\beta,\gamma)\nonumber\\
\end{eqnarray}

where $E=0.5624Mhz$ and $D=4.4450Mhz$, $(g_x,g_y,g_z)=
(2.86,3.05,11.56)kHz/G$, and the Euler angles are
$(\alpha,\beta,\gamma)=(-99.7,55.7,-40)$ \cite{rotpaper}.  These
values are for the crystal aligned with the $C_2$ axis in the $y$
direction, and the $z$ axis is the direction of linear polarization of
the \Pr\ optical transitions.  These tensors are highly anisotropic
due to the low symmetry of the site.

The dominant dephasing mechanism for the Pr hyperfine ground states in
\YSO\ is due to magnetic interactions with the Y nuclear spins in the
host. Yttrium has a nuclear spin of 1/2 and a gyromagnetic ratio of
$g_n=200Hz/G$. Nearest neighbor Y ions induce a magnetic field at the
Pr site of the order of 0.1G.  The direct magnetic dipole-dipole
interaction between the Pr and a nearest neighbor Y ion is of the
order of hundreds of Hertz. This interaction can induce decoherence
via two mechanisms.  The first mechanism is that the Y ions can
exchange spin with each other, resulting in a fluctuating magnetic
field at the Pr site. For resonant Y nuclei the rate of exchange
between two nearest neighbor ions is of the order of tens of Hertz.
The second mechanism involves the excitation of nearly degenerate
transitions where one or more Y spin flips are induced along with the
desired change in the Pr hyperfine state.

To suppress these dephasing mechanisms for a given Pr ground state
hyperfine transition we propose applying a specific magnetic field
such that transition's linear Zeeman shift about this field value is
zero.  The main requirements for this technique are the existence of a
zero field splitting in the spin states and the ability to apply a
sufficiently strong magnetic field such that the Zeeman splitting is
comparable to the zero field splittings.  This method should be
applicable to many $I>1$ spin systems.

Using equation \ref{eq:QuadTensor} and the parameters for the site 1
given above, all the magnetic field values resulting in a zero first
Zeeman shift were identified. The case chosen to experimentally
investigate was at a magnetic field of
$\bm{B_{\text{CP}}}=\left\{732,173,-219\right\}G$ on the
$m_I=+1/2\leftrightarrow+3/2$ transition at $8.63$ MHz as it was found
to have the smallest second order Zeeman shift. Around this magnetic
field value the transition energy as a function of magnetic field has
a turning point in the $y$ and $z$ axis while the $x$ axis has a slow
inflection point. As the required magnetic field was not along the
crystal's $C_2$ axis the degeneracy between site 1a and site 1b was
lifted with only the site 1a ions being at the critical point.

\begin{figure}
  \includegraphics{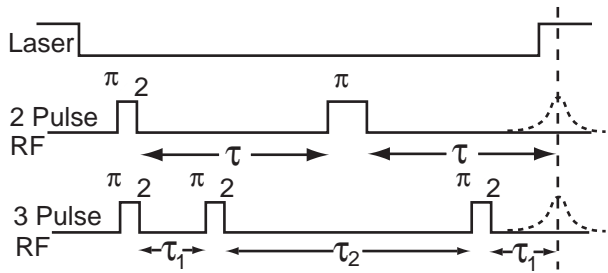}
\caption{\label{fig:pulseSequence} Two and three pulse echo
  sequences. $\pi$ pulse duration was $40\mu\text{s}$ and the laser
  was turned off $150\mu\text{s}$ before the pulse sequence and turned
  on $100\mu\text{s}$ before the echo}
\end{figure}

Raman heterodyne two and three pulse spin echoes were used to
investigate the decoherence and spectral diffusion in the ground state
hyperfine transitions\cite{ham98a} (figure \ref{fig:pulseSequence}).
The experiment was performed using a Coherent 699 frequency stabilized
(1MHz FWHM) tunable dye laser tuned to the ${}^3\!H_{4}-{}^1\!D_2$
transition at 605.977nm. The laser was gated using a 100MHz AOM such
that there was no laser radiation applied to the sample during the RF
pulse sequence (figure \ref{fig:pulseSequence}). The laser power
incident on the crystal was 40mW, focused to $\sim$100$\mu$m and a RF
Rabi frequency of $\Omega_{RF}=300kHz$.  The \PrYSO\ (0.05\%
concentration) crystal was held at $\sim1.2K$ for the duration of the
experiment.  The laser prepared a population difference in the excited
ions for 5s before the pulse sequence and was scanned over 1.2GHz of
the optical transition to avoid hole burning effects.  The laser is
off during the RF pulse sequence to minimize coherence loss from
optical pumping.

\begin{figure}
  \includegraphics[width=0.4\textwidth]{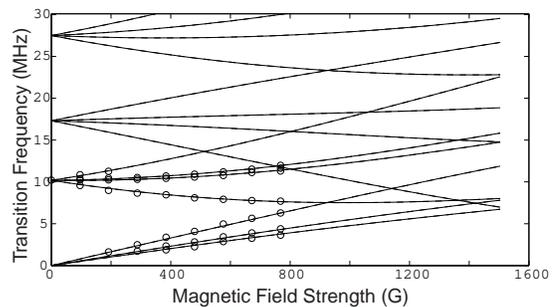}
\caption{\label{fig:ramanhetro} Spectrum of hyperfine transitions
  showing experimentally measured points as the field is increased
  toward the critical point}
\end{figure}

The magnetic fields were supplied by two superconducting magnets one
along the $z$ axis the other in the horizontal plane. The sample was
then rotated by $13\pm0.5^{\circ}$ to provide the correct ratio of
fields along the $x$ and $y$ axes for the critical point in magnetic
field space. Corse adjustment of the field was done by comparing the
Raman heterodyne spectrum to theoretical simulations as shown in
figure \ref{fig:ramanhetro}.  Fine adjustment of the magnetic field
utilized perturbing coils along each axis and associated lock-in
amplifiers to perform field sensitivity measurements on the frequency
of the $m_I=+1/2\leftrightarrow+3/2$ transition.  The rotation of the
sample and the current driving the magnets were iteratively adjusted
to minimize the sensitivity of the desired transition.  Final
adjustments of the field values were made by minimizing the echo decay
rate.  The magnetic field values were accurate to within $\pm$$0.1G$
of the critical point.

\begin{figure}
  \includegraphics[width=0.45\textwidth]{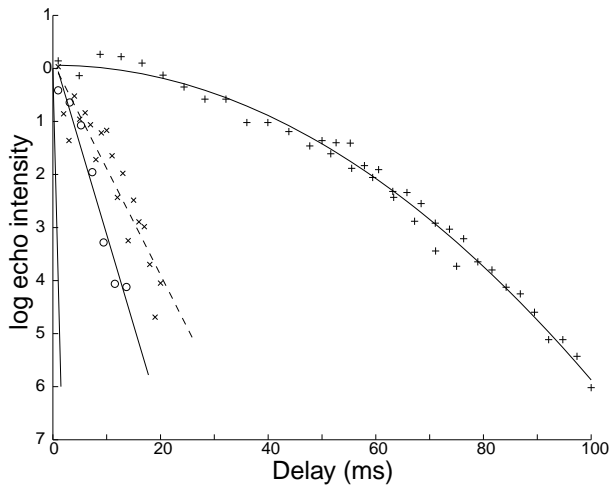}
\caption{\label{fig:2pulseEcho} Two pulse echo sequences taken at the
  critical point magnetic field
  $\bm{B_{\text{CP}}}=\left\{732,172,-219\right\}G$ for transition
  $m_I=+1/2\leftrightarrow+3/2$ at site 1a (+). The
  $m_I=+1/2\leftrightarrow+3/2$ site 1b transition ($\times$); the
  $m_I=-3/2\leftrightarrow+3/2$ transition ($\circ$) and the zero
  field $T_2=500\mu\text{s}$ decay line shown for comparison.}
\end{figure}

The two pulse echo data (figure \ref{fig:2pulseEcho}) shows three echo
sequences at the critical point magnetic field configuration with the
zero field echo sequence for reference. Since the field configuration
is transition specific the $m_I=+1/2\leftrightarrow+3/2$ transition at
site 1a is at a critical point while the two other transitions are
not.

With the application of the field for all the transitions there was
observed a significant increase in the coherence time.  For the
$m_I=+3/2\leftrightarrow-3/2$ transition $T_2$ was 5.86ms while for
the $m_I=+1/2\leftrightarrow+3/2$ transition at site 1b it was 9.98ms
(figure \ref{fig:2pulseEcho}).  Applying a field lifts the degeneracy
of the Y spin states and inhibits Pr transition involving single Y
spin flips\cite{sell95}.  This can be understood by considering the Y
quantization axis.  At zero field the Y quantization axis is locally
defined by the Pr ion. Therefore a change in the spin state changes
the quantization axis, mixing the Y spin states and resulting in a
high spin flip probability.  With an applied field, large compared to
the Pr-Y coupling, changes in the Pr spin state only weakly perturb
the quantization axis.

The echo decay for the $m_I=+1/2\leftrightarrow+3/2$ at site 1a is
significantly slower than for the other two transitions and is clearly
not described by a simple exponential.  Fitted to this decay is an
exponential with a quadratic time dependence, equation
\ref{eq:PhaseMemory}.  This coherence decay is indicative of a
spectral diffusion process described by a Lorentz kernel with a width
increasing linearly with time at a rate of $47.3 Hz/s$\cite{mims68}.
The phase memory, $T_M$ is used to characterise the decoherence rate
of the system.  In the present case the $T_M$ was determined to be 82
ms.

\begin{equation}
  \label{eq:PhaseMemory}
  I(2t)=I_{0}exp\left[{-\left(\frac{2t}{T_M}\right)^2}\right]
\end{equation}

\begin{figure}
\begin{center}
$\begin{array}{cr}
  {\bf (a)} &
  \includegraphics[width=0.455\textwidth]{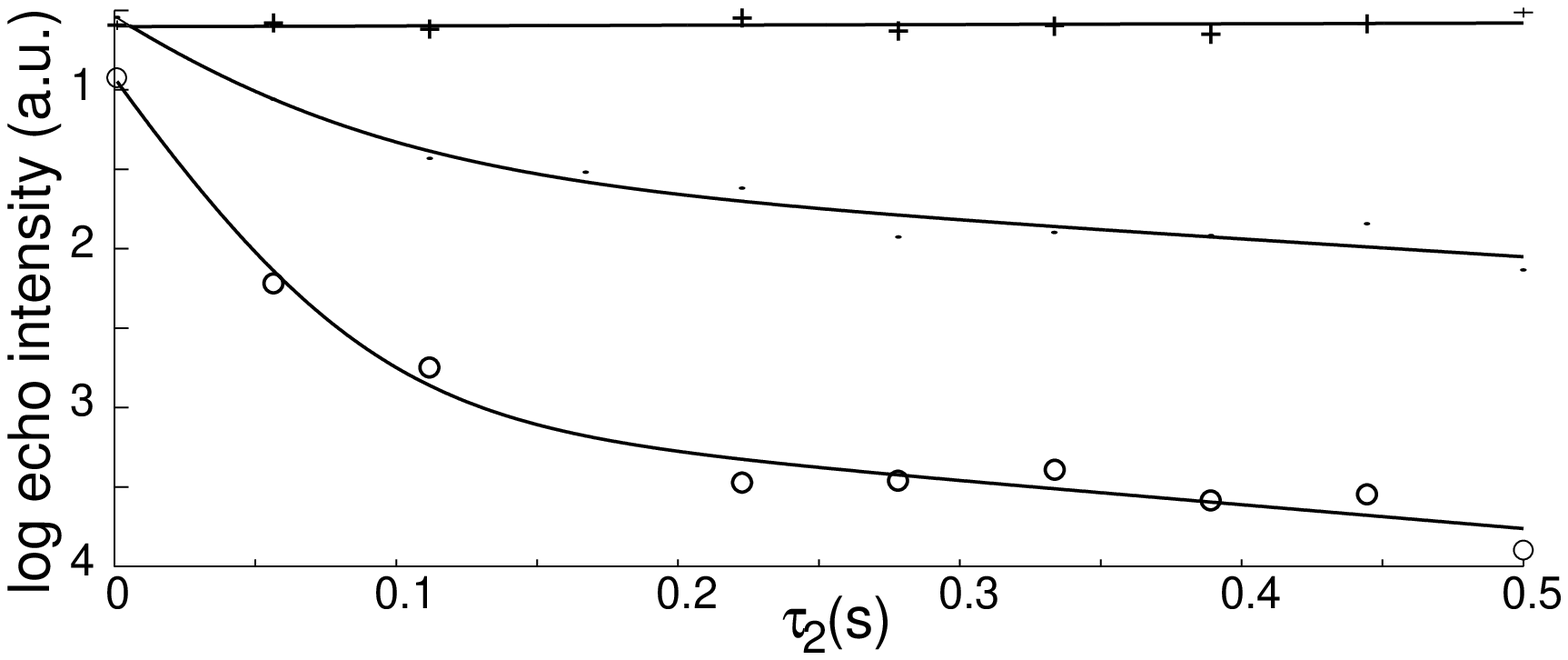}\\
  {\bf (b)} &
  \includegraphics[width=0.455\textwidth]{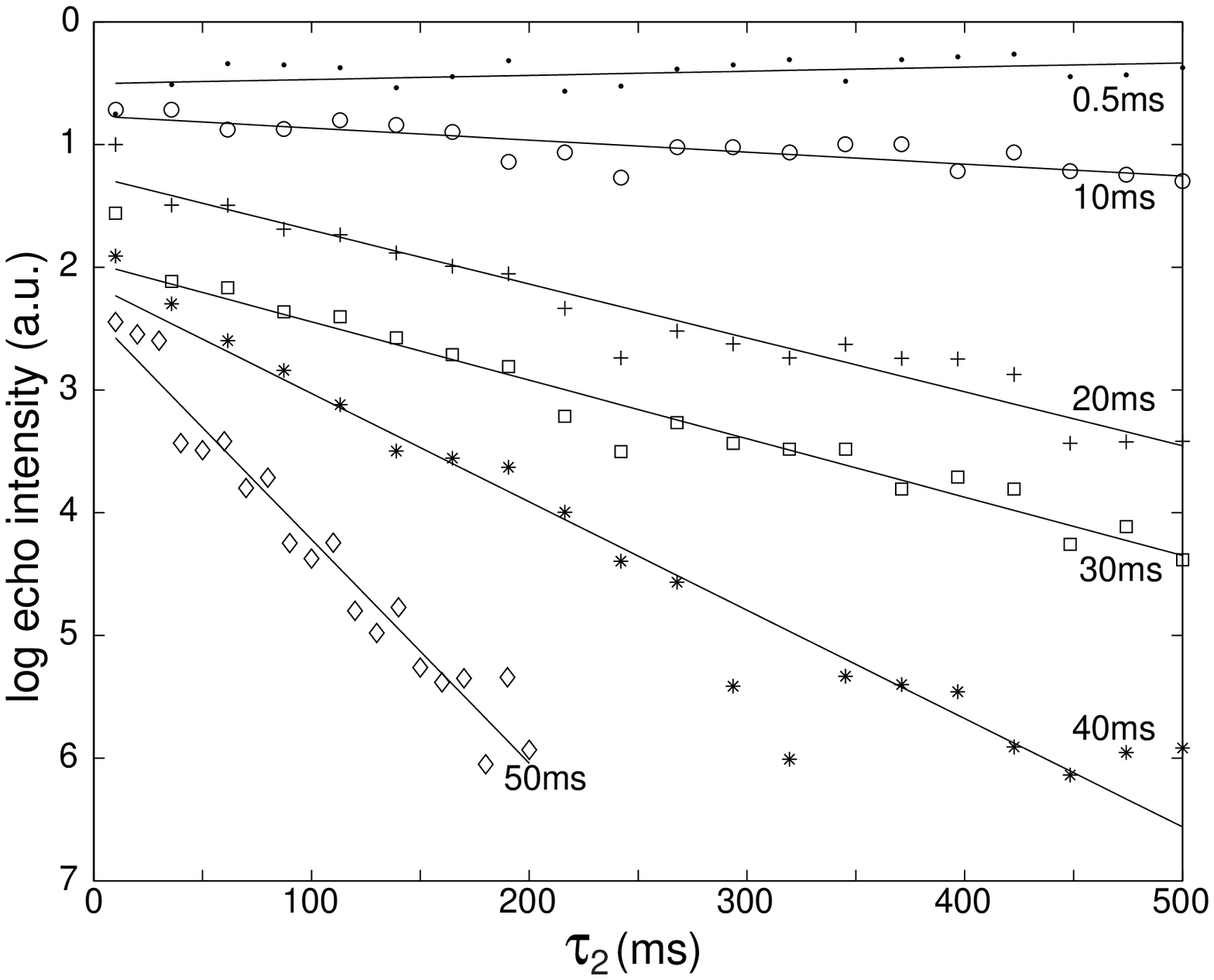}\\
\end{array}$
\end{center}
\caption{Three pulse echo sequences taken at (a) $m_I=+1/2\leftrightarrow+3/2$ transition at site 1a(+) and site 1b(.)  $m_I=-3/2\leftrightarrow+3/2$ transition ($\circ$), both traces were taken with $\tau_1=2.5ms$. The delay $\tau_2$ is measured on the $x$ axis, (b) Critical point on transition  $m_I=+1/2\leftrightarrow+3/2$ at site 1a with $\tau_1$ delays are labeled on the fit line }
\label{fig:3pulseEchos}
\end{figure}

To investigate the diffusion process over longer times scales a series
of three pulse echo measurements were made. Within each of the series
shown in Fig \ref{fig:3pulseEchos} the delay between the first two
pulses ($\tau_1$) is held constant while the delay between the second
and third pulses ($\tau_2$) is varied.

A comparison is made with the same transitions as used in the two
pulse echo study.  Comparing echo sequences with the same $\tau_1$ on
the different transitions shows a clear difference between the
critical point and the other two transitions (figure
\ref{fig:3pulseEchos}a).  The $m_I=-3/2\leftrightarrow+3/2$ and the
site 1b $m_I=+1/2\leftrightarrow+3/2$ transitions have a clearly
biexponential decay: a fast initial decay, reducing to a slow decay
for $\tau_2>200ms$.  With a $\tau_1$ of 2.5ms no spectral diffusion
was observed on the critical point transition.

The biexponential decay is attributed to the existence of a ``frozen
core''\cite{szab83,wald92}.  A frozen core is formed by the Pr ion
detuning nearby Y ions from the bulk Y, thereby inhibiting mutual Y
spin flips within the frozen core.  A strong frozen core is expected
due to the Pr magnetic moment ratio being an order of magnitude larger
than that of Y.  Although a frozen core has not been reported in
\PrYSO, it has been observed in directly analagous
systems\cite{wald92}.  On short time scales the rapid spin flips of
the bulk Y ions dominate the dephasing.  Since the bulk Y are well
removed from the Pr ion they only weakly perturb the transition
frequency.  On longer time scales the frozen core Y spin flips
dominate the dephasing as they induce larger but less frequent
frequency shifts.  Given that the slow spectral diffusion has not
plateaued for the longest $\tau_2$ used, the correlation time of the
frozen core is comparable or longer than 0.5s.

Further investigation of the three pulse echos on the critical point
transition (figure \ref{fig:3pulseEchos}) shows that for all values of
$\tau_1$ only a single, slow exponential decay was observed.
Therefore dephasing of the critical point transition dominated, at all
time scales by the frozen core Y spin flips.  The fact that the bulk Y
dephasing contribution is absent for the site 1a
$m_I=+1/2\leftrightarrow+3/2$ transition and present for the other two
transitions is consistent with the dephasing at the critical point
being due to second order Zeeman interactions.  This suggests that the
applied field is accurate to the order of a Pr-Y interaction strength
from the ideal critical point.  Further optimisation of the applied
field is therefore unlikely to produce a significant gain in coherence
time.

Using the method described above a regime has reached where the
correlation time of the dephasing interaction is extremely long
compared to the duration of readily achievable RF pulses.  This is
exactly the regime required for effective implementation of quantum
error correction schemes\cite{viol98,vital99,byrd02}.  The simplest
implementation shown to be sufficient for error correction in solid
state qbits\cite{viol98} consists of a series of hard rephasing pulses
($\pi$) separated by the cycling time.  In the present case the
difference between a transition specific hard RF $\pi$ pulse ($\geq
1\mu s$) and the $T_M$ would allow an effective error correction
scheme to be applied with a duty cycle $<0.1\%$.

We have shown that a phase memory time of 82ms can be achieved for
\PrYSO\ by using an external magnetic field to minimise the transition
sensitivity to magnetic field fluctuations.  Directions to further
reduce the residual dephasing mechanism have been identified.  We have
shown that even in hosts containing nuclear spins it is possible to
obtain spin based qbits that have coherence properties suitable for
sophisticated quantum computing demonstrations.

\bibliography{turningpoint.bib}

\begin{thebibliography}{17}
\expandafter\ifx\csname natexlab\endcsname\relax\def\natexlab#1{#1}\fi
\expandafter\ifx\csname bibnamefont\endcsname\relax
  \def\bibnamefont#1{#1}\fi
\expandafter\ifx\csname bibfnamefont\endcsname\relax
  \def\bibfnamefont#1{#1}\fi
\expandafter\ifx\csname citenamefont\endcsname\relax
  \def\citenamefont#1{#1}\fi
\expandafter\ifx\csname url\endcsname\relax
  \def\url#1{\texttt{#1}}\fi
\expandafter\ifx\csname urlprefix\endcsname\relax\def\urlprefix{URL }\fi
\providecommand{\bibinfo}[2]{#2}
\providecommand{\eprint}[2][]{\url{#2}}

\bibitem[{\citenamefont{Ichimura}(2001)}]{ichi01}
\bibinfo{author}{\bibfnamefont{K.}~\bibnamefont{Ichimura}},
  \bibinfo{journal}{Opt. Commun.} \textbf{\bibinfo{volume}{196}},
  \bibinfo{pages}{119} (\bibinfo{year}{2001}).

\bibitem[{\citenamefont{Kane}(1998)}]{kane98}
\bibinfo{author}{\bibfnamefont{B.}~\bibnamefont{Kane}},
  \bibinfo{journal}{Nature} \textbf{\bibinfo{volume}{393}},
  \bibinfo{pages}{133} (\bibinfo{year}{1998}).

\bibitem[{\citenamefont{Ohlsson et~al.}(2002)\citenamefont{Ohlsson, Mohan, and
  Kr\"oll}}]{ohls02}
\bibinfo{author}{\bibfnamefont{N.}~\bibnamefont{Ohlsson}},
  \bibinfo{author}{\bibfnamefont{R.~K.} \bibnamefont{Mohan}}, \bibnamefont{and}
  \bibinfo{author}{\bibfnamefont{S.}~\bibnamefont{Kr\"oll}},
  \bibinfo{journal}{Optics Comm.} \textbf{\bibinfo{volume}{201}},
  \bibinfo{pages}{71} (\bibinfo{year}{2002}).

\bibitem[{\citenamefont{Turukin et~al.}(2002)\citenamefont{Turukin,
  Sudarshanam, Shahrier, Musser, Ham, and Hemmer}}]{turu02}
\bibinfo{author}{\bibfnamefont{A.~V.} \bibnamefont{Turukin}},
  \bibinfo{author}{\bibfnamefont{V.~S.} \bibnamefont{Sudarshanam}},
  \bibinfo{author}{\bibfnamefont{M.~S.} \bibnamefont{Shahrier}},
  \bibinfo{author}{\bibfnamefont{J.~A.} \bibnamefont{Musser}},
  \bibinfo{author}{\bibfnamefont{B.~S.} \bibnamefont{Ham}}, \bibnamefont{and}
  \bibinfo{author}{\bibfnamefont{P.~R.} \bibnamefont{Hemmer}},
  \bibinfo{journal}{Phys.\ Rev.\ Lett.} \textbf{\bibinfo{volume}{88}},
  \bibinfo{pages}{023602} (\bibinfo{year}{2002}).

\bibitem[{\citenamefont{Yamamoto et~al.}(1998)\citenamefont{Yamamoto, Ichimura,
  and Gemma}}]{yama98}
\bibinfo{author}{\bibfnamefont{K.}~\bibnamefont{Yamamoto}},
  \bibinfo{author}{\bibfnamefont{K.}~\bibnamefont{Ichimura}}, \bibnamefont{and}
  \bibinfo{author}{\bibfnamefont{N.}~\bibnamefont{Gemma}},
  \bibinfo{journal}{Phys.\ Rev.\ A} \textbf{\bibinfo{volume}{58}},
  \bibinfo{pages}{2461} (\bibinfo{year}{1998}).

\bibitem[{\citenamefont{Ham et~al.}(1998)\citenamefont{Ham, Shahriar, and
  Hemmer}}]{ham98a}
\bibinfo{author}{\bibfnamefont{B.~S.} \bibnamefont{Ham}},
  \bibinfo{author}{\bibfnamefont{M.~S.} \bibnamefont{Shahriar}},
  \bibnamefont{and} \bibinfo{author}{\bibfnamefont{P.~R.}
  \bibnamefont{Hemmer}}, \bibinfo{journal}{Phys.\ Rev.\ B}
  \textbf{\bibinfo{volume}{58}}, \bibinfo{pages}{R11825}
  (\bibinfo{year}{1998}).

\bibitem[{\citenamefont{Buerger}(1969)}]{buer69}
\bibinfo{author}{\bibfnamefont{M.~J.} \bibnamefont{Buerger}},
  \emph{\bibinfo{title}{International Tables For X-Ray Crystallography}}
  (\bibinfo{publisher}{Kynoch Press}, \bibinfo{address}{Birmingham, England},
  \bibinfo{year}{1969}), \bibinfo{edition}{3rd} ed.

\bibitem[{\citenamefont{Equall et~al.}(1995)\citenamefont{Equall, Cone, and
  Macfarlane}}]{equa95}
\bibinfo{author}{\bibfnamefont{R.~W.} \bibnamefont{Equall}},
  \bibinfo{author}{\bibfnamefont{R.~L.} \bibnamefont{Cone}}, \bibnamefont{and}
  \bibinfo{author}{\bibfnamefont{R.~M.} \bibnamefont{Macfarlane}},
  \bibinfo{journal}{Phys.\ Rev.\ B.} \textbf{\bibinfo{volume}{52}},
  \bibinfo{pages}{3963} (\bibinfo{year}{1995}).

\bibitem[{\citenamefont{Baker and Bleaney}(1958)}]{bake58}
\bibinfo{author}{\bibfnamefont{J.~M.} \bibnamefont{Baker}} \bibnamefont{and}
  \bibinfo{author}{\bibfnamefont{B.}~\bibnamefont{Bleaney}},
  \bibinfo{journal}{Proc. R. Soc. London Ser. A}
  \textbf{\bibinfo{volume}{245}}, \bibinfo{pages}{156} (\bibinfo{year}{1958}).

\bibitem[{\citenamefont{Longdell et~al.}(2002)\citenamefont{Longdell, Sellars,
  and Manson}}]{rotpaper}
\bibinfo{author}{\bibfnamefont{J.~J.} \bibnamefont{Longdell}},
  \bibinfo{author}{\bibfnamefont{M.~J.} \bibnamefont{Sellars}},
  \bibnamefont{and} \bibinfo{author}{\bibfnamefont{N.~B.}
  \bibnamefont{Manson}}, \bibinfo{journal}{Phys.\ Rev.\ B}
  \textbf{\bibinfo{volume}{66}}, \bibinfo{pages}{035101}
  (\bibinfo{year}{2002}).

\bibitem[{\citenamefont{Sellars}(1995)}]{sell95}
\bibinfo{author}{\bibfnamefont{M.~J.} \bibnamefont{Sellars}}, Ph.D. thesis,
  \bibinfo{school}{RSPhySSE, ANU} (\bibinfo{year}{1995}).

\bibitem[{\citenamefont{Mims}(1968)}]{mims68}
\bibinfo{author}{\bibfnamefont{W.~B.} \bibnamefont{Mims}},
  \bibinfo{journal}{Phys.\ Rev.} \textbf{\bibinfo{volume}{168}},
  \bibinfo{pages}{370} (\bibinfo{year}{1968}).

\bibitem[{\citenamefont{Wald et~al.}(1992)\citenamefont{Wald, Hahn, and
  Lunac}}]{wald92}
\bibinfo{author}{\bibfnamefont{L.~L.} \bibnamefont{Wald}},
  \bibinfo{author}{\bibfnamefont{E.~L.} \bibnamefont{Hahn}}, \bibnamefont{and}
  \bibinfo{author}{\bibfnamefont{M.}~\bibnamefont{Lunac}},
  \bibinfo{journal}{J.\ Opt.\ Soc.\ Am.\ B} \textbf{\bibinfo{volume}{9}},
  \bibinfo{pages}{789} (\bibinfo{year}{1992}).

\bibitem[{\citenamefont{Szabo}(1983)}]{szab83}
\bibinfo{author}{\bibfnamefont{A.}~\bibnamefont{Szabo}}, \bibinfo{journal}{Opt.
  Lett.} \textbf{\bibinfo{volume}{8}}, \bibinfo{pages}{486}
  (\bibinfo{year}{1983}).

\bibitem[{\citenamefont{Viola and Lloyd}(1998)}]{viol98}
\bibinfo{author}{\bibfnamefont{L.}~\bibnamefont{Viola}} \bibnamefont{and}
  \bibinfo{author}{\bibfnamefont{S.}~\bibnamefont{Lloyd}},
  \bibinfo{journal}{Phys.\ Rev.\ A.} \textbf{\bibinfo{volume}{58}},
  \bibinfo{pages}{2733} (\bibinfo{year}{1998}).

\bibitem[{\citenamefont{Vitali and Tombesi}(1999)}]{vital99}
\bibinfo{author}{\bibfnamefont{D.}~\bibnamefont{Vitali}} \bibnamefont{and}
  \bibinfo{author}{\bibfnamefont{P.}~\bibnamefont{Tombesi}},
  \bibinfo{journal}{Phys.\ Rev.\ Lett.} \textbf{\bibinfo{volume}{59}},
  \bibinfo{pages}{4178} (\bibinfo{year}{1999}).

\bibitem[{\citenamefont{Byrd and Lidar}(2002)}]{byrd02}
\bibinfo{author}{\bibfnamefont{M.~S.} \bibnamefont{Byrd}} \bibnamefont{and}
  \bibinfo{author}{\bibfnamefont{D.~A.} \bibnamefont{Lidar}},
  \bibinfo{journal}{Phys.\ Rev.\ Lett.} \textbf{\bibinfo{volume}{89}},
  \bibinfo{pages}{047901} (\bibinfo{year}{2002}).

\end{thebibliography}

\end{document}